%% file: soft_Matter_V4.tex
\documentclass[twoside,twocolumn,9pt]{article}
\usepackage{extsizes}
\usepackage[super,sort&compress,comma]{natbib} 
\usepackage[version=3]{mhchem}
\usepackage[left=1.5cm, right=1.5cm, top=1.785cm, bottom=2.0cm]{geometry}
\usepackage{balance}
\usepackage{mathptmx}
\usepackage{sectsty}
\usepackage{graphicx} 
\usepackage{bm}
\usepackage{lastpage}
\usepackage[format=plain,justification=justified,singlelinecheck=false,font={stretch=1.125,small,sf},labelfont=bf,labelsep=space]{caption}
\usepackage{float}
\usepackage{fancyhdr}
\usepackage{fnpos}
\usepackage[english]{babel}
\addto{\captionsenglish}{%
  
}
\usepackage{array}
\usepackage{droidsans}
\usepackage{charter}
\usepackage[T1]{fontenc}
\usepackage[usenames,dvipsnames]{xcolor}
\usepackage{setspace}
\usepackage[compact]{titlesec}
\usepackage{hyperref}

\usepackage{latexsym}
\usepackage{amsmath}
\usepackage{amsfonts}
\usepackage{amssymb}
\usepackage{amsbsy}
\usepackage{wasysym}
\usepackage{marvosym}
\usepackage{oplotsymbl}

\newcommand{\cblue }{\color{black}}

\usepackage{epstopdf}

\definecolor{cream}{RGB}{222,217,201}

\begin{document}

\pagestyle{fancy}
\thispagestyle{plain}
\fancypagestyle{plain}{
\renewcommand{\headrulewidth}{0pt}
}

\makeFNbottom
\makeatletter
\renewcommand\LARGE{\@setfontsize\LARGE{15pt}{17}}
\renewcommand\Large{\@setfontsize\Large{12pt}{14}}
\renewcommand\large{\@setfontsize\large{10pt}{12}}
\renewcommand\footnotesize{\@setfontsize\footnotesize{7pt}{10}}
\makeatother

\renewcommand{\thefootnote}{\fnsymbol{footnote}}
\renewcommand\footnoterule{\vspace*{1pt}%
\color{cream}\hrule width 3.5in height 0.4pt \color{black}\vspace*{5pt}} 
\setcounter{secnumdepth}{5}

\makeatletter 
\renewcommand\@biblabel[1]{#1}            
\renewcommand\@makefntext[1]%
{\noindent\makebox[0pt][r]{\@thefnmark\,}#1}
\makeatother 
\renewcommand{\figurename}{\small{Fig.}~}
\sectionfont{\sffamily\Large}
\subsectionfont{\normalsize}
\subsubsectionfont{\bf}
\setstretch{1.125} 
\setlength{\skip\footins}{0.8cm}
\setlength{\footnotesep}{0.25cm}
\setlength{\jot}{10pt}
\titlespacing*{\section}{0pt}{4pt}{4pt}
\titlespacing*{\subsection}{0pt}{15pt}{1pt}

\fancyfoot{}
\fancyfoot[LO,RE]{\vspace{-7.1pt}\includegraphics[height=9pt]{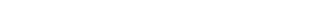}}
\fancyfoot[CO]{\vspace{-7.1pt}\hspace{13.2cm}\includegraphics{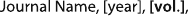}}
\fancyfoot[CE]{\vspace{-7.2pt}\hspace{-14.2cm}\includegraphics{RF}}
\fancyfoot[RO]{\footnotesize{\sffamily{1--\pageref{LastPage} ~\textbar  \hspace{2pt}\thepage}}}
\fancyfoot[LE]{\footnotesize{\sffamily{\thepage~\textbar\hspace{3.45cm} 1--\pageref{LastPage}}}}
\fancyhead{}
\renewcommand{\headrulewidth}{0pt} 
\renewcommand{\footrulewidth}{0pt}
\setlength{\arrayrulewidth}{1pt}
\setlength{\columnsep}{6.5mm}
\setlength\bibsep{1pt}

\makeatletter 
\newlength{\figrulesep} 
\setlength{\figrulesep}{0.5\textfloatsep} 

\newcommand{\topfigrule}{\vspace*{-1pt}%
\noindent{\color{cream}\rule[-\figrulesep]{\columnwidth}{1.5pt}} }

\newcommand{\botfigrule}{\vspace*{-2pt}%
\noindent{\color{cream}\rule[\figrulesep]{\columnwidth}{1.5pt}} }

\newcommand{\dblfigrule}{\vspace*{-1pt}%
\noindent{\color{cream}\rule[-\figrulesep]{\textwidth}{1.5pt}} }

\makeatother

\twocolumn[
  \begin{@twocolumnfalse}
{\includegraphics[height=30pt]{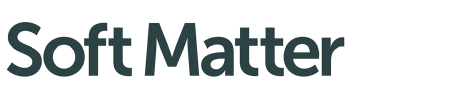}\hfill\raisebox{0pt}[0pt][0pt]{\includegraphics[height=55pt]{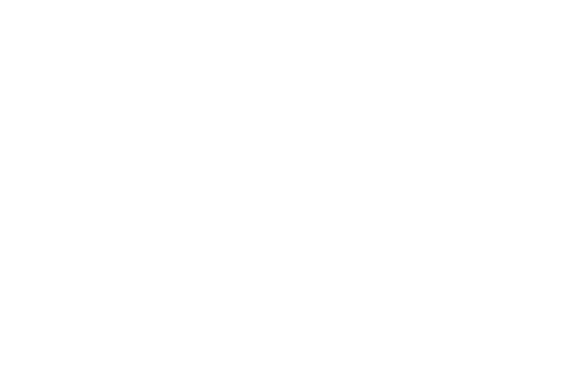}}\\[1ex]
\includegraphics[width=18.5cm]{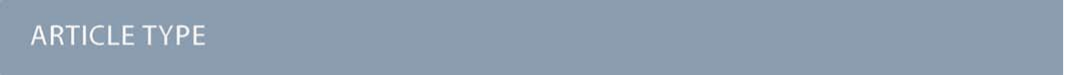}}\par
\vspace{1em}
\sffamily
\begin{tabular}{m{4.5cm} p{13.5cm} }

\includegraphics{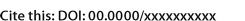} & \noindent\LARGE{\textbf{Characteristic  features of self-avoiding active Brownian polymers under linear shear flow$^\dag$}} \\
\vspace{0.3cm} & \vspace{0.3cm} \\

 & \noindent\large{Arindam Panda,$^{\ast}$\textit{$^{a}$}, Roland G. Winkler\textit{$^{b}$}, and Sunil P. Singh,\textit{$^{a\ddag}$}} \\

\includegraphics{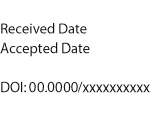} &
\noindent\normalsize{ 
We present Brownian dynamics simulation results of a flexible linear polymer with excluded-volume interactions under shear flow in the presence of active noise. The active noise strongly affects the polymer's conformational and dynamical properties, such as the stretching in the flow direction and compression in the gradient direction, shear-induced alignment, and shear viscosity.  In the asymptotic limit of large activities and shear rates, the power-law scaling exponents of these quantities differ significantly from those of passive polymers. The chain's shear-induced stretching at a given shear rate is reduced by active noise, and it displays a non-monotonic behavior, where an initial polymer compression is followed by its stretching with increasing active force. 
The compression of the polymer in the gradient direction follows the relation $\sim Wi_{Pe}^{-3/4}$ as a function of the activity-dependent Weissenberg number $Wi_{Pe}$, which differs from the scaling observed in passive systems $\sim Wi_{Pe}^{-1/2}$. The flow-induced alignment at large P\'eclet numbers  $Pe \gg 1$, where $Pe$ is the P\'eclet number, and large shear rates $Wi_{Pe} \gg 1$ displays the scaling behavior $Wi_{Pe}^{-1/2}$, with an exponent differing from the passive value  $-1/3$. 
Furthermore, the polymer's zero-shear viscosity displays a non-monotonic behavior, decreasing in an intermediate activity regime due to excluded-volume interactions and increasing again for large $Pe$.
Shear thinning appears with increasing Weissenberg number with the power-laws  $Wi_{Pe}^{-1/2}$ and $Wi_{Pe}^{-3/4}$ for passive and active polymers, respectively. In addition, our simulation results are compared with the results of an analytical approach, which predicts quantitatively similar behaviors for the various aforementioned physical quantities. }



\\

\end{tabular}

 \end{@twocolumnfalse} \vspace{0.6cm}
]
  

\renewcommand*\rmdefault{bch}\normalfont\upshape
\rmfamily
\section*{}
\vspace{-1cm}


\footnotetext{\textit{$^{a}$~Department of Physics, Indian Institute Of Science Education and Research, Bhopal 462 066, Madhya Pradesh, India; E-mail: arindam19@iiserb.ac.in }}
\footnotetext{\textit{$^{b}$~Theoretical Physics of Living Matter, Institute of Biological Information Processing and Institute for Advanced Simulation, Forschungszentrum J{\"u}lich, 52428 J{\"u}lich, Germany; E-mail: r.winkler@fz-juelich.de}}
\footnotetext{\textit{$^{a\ddag}$~Department of Physics, Indian Institute Of Science Education and Research, Bhopal 462 066, Madhya Pradesh, India; E-mail: spsingh@iiserb.ac.in }}



\section{Introduction}

 Biological systems comprise a plethora of filaments, polymers, and polymer-like structures, and their conformational and dynamical properties are either determined by active processes, e.g., due to self-propulsion, or substantially affected by active processes to a yet unresolved extent.\cite{winkler2020physics}
Bio-polymers, such as actin filaments, microtubules, RNA, and various proteins, often encounter athermal active noise.\cite{ramaswamy2010mechanics,marchetti2013hydrodynamics} 
The active noise originates from the environment, such as stresses generated by the directional motion of molecular motors, which harnesses chemical energy from the hydrolysis of the ATP to ADP conversion.\cite{winkler2017active,schaller2010polar,prost2015active,doostmohammadi2017onset,ndlec1997self,anand2019behavior,manna2019emergent,ganguly2012cytoplasmic,harada1987sliding,juelicher2007active,foglino2019non,kapral2013perspective,kapral2016stirring} These local stresses drive the system out of equilibrium and instigates several interesting dynamical processes at various temporal scales viz. enhanced local diffusion,\cite{anand2018structure,bianco2018globulelike,samanta2016chain,eisenstecken2016conformational,eisenstecken2017internal,winkler2020physics} transport of vesicles,\cite{ahmed2014active,raote2019protein,goode2000functional,schuh2011actin,kamal2000connecting} cell migration,\cite{horwitz2003cell,trepat2012cell,lauffenburger1996cell,yamada2019mechanisms,friedl2012new} etc. Additionally, such active stresses induce bending and buckling,\cite{chelakkot2014flagellar,laskar2013hydrodynamic,anand2019beating, anand2018structure} large-scale conformational deformations, \cite{anand2018structure,bianco2018globulelike,samanta2016chain,eisenstecken2016conformational,eisenstecken2017internal,osmanovic2017dynamics,martin2018active,shin2015facilitation,paul2022activity} and bundling of bio-polymers.\cite{ramaswamy2010mechanics,ndlec1997self,juelicher2007active,schaller2010polar,sanchez2012spontaneous} 

Biological microswimmers, like algae, sperm, and bacteria, are omnipresent.\cite{berg:04,elge:15,rama:10} Many of them are elongated and polymer- or filament-like, undergoing shape changes during migration. Swarming bacteria, propelled by flagella, such as Proteus mirabilis, Vibrio parahaemolyticus,\cite{weib:14,auer:19} and Serratia marcescens,\cite{li:19} are particular examples. Moreover, certain microswimmers organize in chain-like structures, e.g., planktonic dinoflagellates\cite{sohn:11,sela:11} or {\em Bacillus subtilis} bacteria during biofilm formation.\cite{yama:19}  Dynamical networks are formed by nematodes such as {\em Caenorhabditis elegans} on a more macroscopic scale. \cite{gagn:16,sugi:19} 

Various routes have been proposed to synthesize active or activated colloidal molecules\cite{lowen2018active} or polymers.\cite{winkler2017active} Often, phoretic effects are exploited for propulsion by local gradients of, e.g.,  electric fields (electrophoresis), concentration (diffusiophoresis),  and temperature (thermophoresis).\cite{hows:07,jian:10,vala:10,wurg:10,volp:11,thut:11,butt:13,hage:14,bech:16,maas:16}
Metal–dielectric Janus colloids assemble into active chains by imbalanced interactions, and motility as well as colloidal interactions are simultaneously controlled by an AC electric field.\cite{yan:16,dile:16,zhan:16,zhan:16.1,nish:18,mart:19} Strong AC electric fields induce the formation of linear chains of bound Janus particles either by van-der-Waals forces or polymer linkers.\cite{vutu:17} Self-assembling of colloidal chains in a nematic liquid crystal matrix and directed movement is achieved by electrohydrodynamic convection rolls,\cite{sasa:14} and dielectric colloidal particles form chains by alternating magnetic fields.\cite{mart:15,koko:17} Recently, chains of coated catalytic nanoparticles have been assembled,\cite{bisw:17} as well as freely-jointed chains of emulsion droplets moving autonomously.\cite{kumar2023emergent}

Fluid flow is an omnipresent feature of microbial habitats and has strong implications on microbial ecology and environmental microbiology.\cite{rusconi2015flow}
Microorganisms, in particular marine and aquatic microorganisms, frequently encounter flows, which can impact elementary processes such as motility, chemotaxis, and nutrient uptake.\cite{rusconi2015flow}  It has been established that fluid flow plays a decisive role in migration and accumulation of microswimmers at surfaces.\cite{rothschild1963non,woolley2003motility,kaya2012direct,lauga2006swimming,hill2007hydrodynamic,si2020self,chelakkot2010migration,chelakkot2012flow,steinhauser2012mobility} The presence of fluid-flow and active forces lead to intriguing phenomena like upstream swimming in narrow channels\cite{zottl2020simulation,nili2017population,chilukuri2014impact,qi2020rheo} and migration away from a surface.\cite{anand2019behavior,qi2020rheo} Moreover, activity leads to substantial changes in the suspension viscosity under shear flow, as demonstrated for bacteria. \cite{lopez2015turning} The latter aspect renders active polymer-like structures an interesting viscosity modifier. 

With advances in the design of synthetic active and activated polymer-like structures, the application of such molecules in processing may become feasible. Characterization of the emergent properties of such polymers under flow will be helpful in this endeavor.  

In this article, we investigate the conformational and rheological properties of an active Brownian polymer (ABPO) under linear shear flow. In the ABPO model, the active force corresponds to a Gaussian colored noise on a monomer, which changes its direction in a diffusive manner, and its conformational and dynamical properties have been extensively studied in the past few years. \cite{kaiser2014unusual,ghosh2014dynamics,eisenstecken2016conformational,winkler2020physics,eisenstecken2017internal,martin2018active,samanta2016chain,anand2020conformation,paul2022activity} {\cblue Besides that active polymers with Gaussian, but non-local stochastic excitations have been considered for polymer folding and genome organization.\cite{osmanovic2017dynamics,goychuk2023polymer}}

Simulations of ABPOs have already emphasized the significance of excluded-volume interactions, where it was shown that they swell in the absence of excluded-volume interactions\cite{eisenstecken2016conformational,eisenstecken2017conformational,eisenstecken2017internal,winkler2020physics} whereas, in the presence of excluded-volume interactions, they display a weak compression after that swelling prevails in the large activity limit.\cite{anand2020conformation,das2021coil} Notably,  a recent study has corroborated that a passive polymer in an active environment displayed enhanced diffusion in the intermediate regime and non-monotonic structural response as a function of the P\'eclet number\cite{mousavi2021active}, similar to that reported for an ABPO. 


The theoretical study in Ref.\cite{martin2018active} sheds light on the rheological characteristics of dry phantom flexible and semiflexible ABPOs. 
The active polymer's shear behavior is distinct in the extensive activity regime. In particular, the shear viscosity of flexible polymers decreases with increasing Weissenberg number with an exponent ${\beta =1}$ at large activities rather than the value  ${\beta = 1/2}$  of the passive polymer.\cite{martin2018active} {\cblue On the contrary, in experiments on worms,\cite{deblais2020rheology} the shear-thinning exponent for the active filaments is reported to be smaller than unity ($\beta < 1 $), which is believed to be a consequence of the distinct nature of active noise and the concentrated solution in experiments.\cite{deblais2020rheology,martin2018active} }

The current article presents the structural, dynamical,  and rheological characteristics of a flexible ABPO in the linear shear flow. The polymers's conformational changes, i.e., its stretching, alignment, and compression along the flow and gradient direction, are presented. We show that the active noise disrupts the flow-induced alignment of the polymer, and the polymer's end-to-end distance ($R_{e}$) and radius of gyration ($R_{g}$) decrease in the intermediate range of activity, while they swell with the power law $Pe^{1/3}$ in the limit of large P{\'e}clet numbers. A major finding of our simulations is that the polymer's contribution to the zero-shear viscosity decreases in an intermediate regime of activities. We show that this response is different from that of a phantom polymer, where the polymer's contribution monotonically increases with the activity. Furthermore, we show that various physical properties of a polymer attain universal behavior when the simulation results are presented as a function of  Weissenberg number, which is defined with the active noise-dependent relaxation time, such as the compression in gradient direction, shear-induced alignment, etc.  In particular, our simulations highlight the role of the active noise in the large $Pe \gg 1$ regime, where the active polymers's scaling exponents differ significantly from those of passive polymers. 

The article is organized as follows: the next section presents the simulation model (Sec. II). The stretching, compression, alignment, shear thinning behavior, and zero shear viscosity of the active polymer are discussed in various subsections of Sec.~III. The results are summarized in Sec. IV.

 \begin{figure}[t]
	\centering
	\includegraphics[width=\columnwidth]{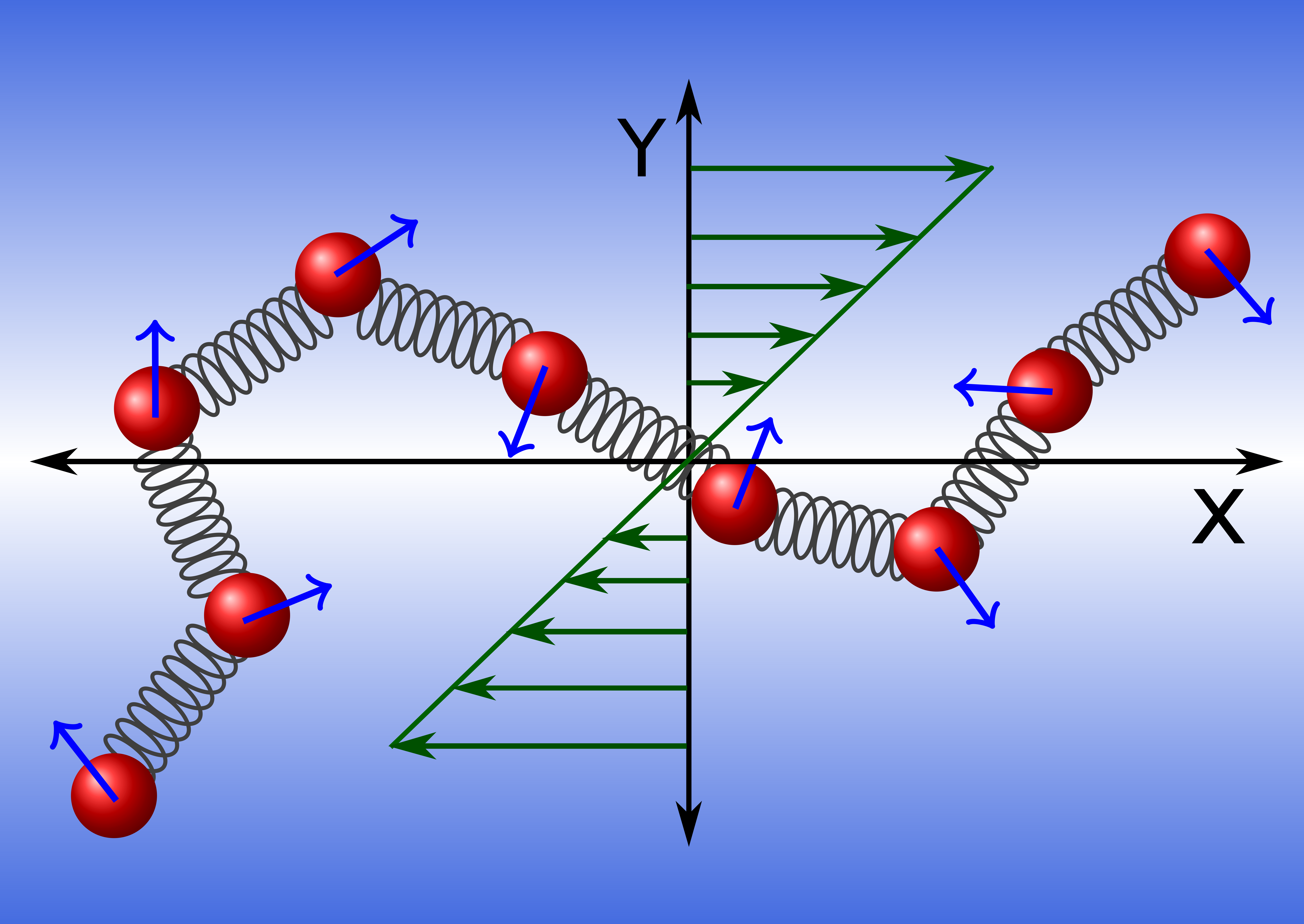}
	\caption{A schematic of an active polymer in the linear shear flow. The arrow on a monomer depicts the direction of the active force. The flow is imposed along the $x$-axis, and the gradient is in the $y$-direction. The color gradient in the background indicates the flow variation in the $y$-direction.  }
	\label{Fig:schematic}
\end{figure}

\section{Model}

The linear polymer is comprised of $N_m$ active monomers connected by the harmonic potential
\begin{equation}
{U_s} = \frac{\kappa_s}{2} \sum_{i=1}^{N_m-1} (|\bm{r}_{i+1} - \bm{r}_i| - l_0)^2,
\label{Eq:spring}
\end{equation} 
where $\kappa_s$ is the force constant, $\bm{r}_i$ denotes the position vector of the $i^{th}$ monomer and $l_0$ is the equilibrium bond length,  which is the smallest length scale in our model. Excluded-volume interactions between non-bonded monomers are captured by the repulsive, truncated, and shifted Lennard-Jones (LJ) potential 
\begin{equation}
	{U_{LJ}} = 4\epsilon \sum_{i>j}^{N_m}  \left[\left(\frac{\sigma}{r_{ij}}\right)^{12} - \left(\frac{\sigma}{r_{ij}}\right)^{6} + \frac{1}{4} \right],
	\label{Eq:LJ}
\end{equation}
for $r_{ij} \leq 2^{1/6} \sigma$, and otherwise $U_{LJ}= 0$.	Here, $\sigma$ is the LJ diameter of the monomers, $\epsilon$ their interaction energy, and ${r}_{ij} = |\bm{r}_j-\bm{r}_i|$ is the distance between monomer $i$ and $j$.\cite{anand2018structure,das2021coil,das2020introduction} Long-range hydrodynamic interactions are ignored, i.e., a dry polymer is considered. The translational equations of motion of the monomers in the over-damped limit are given by 

\begin{equation}
	 \bm{\dot{r}}_i(t) = \frac{1}{\zeta}\left[ \bm{F}_i(t)  + {F}_a \bm{e}_i(t)  +\bm{\Gamma}_i(t)\right] + \dot{\gamma} r_{yi}(t){{\bm e_x}},
	\label{Eq:eq_motion}
\end{equation}
where $\zeta$ is the friction coefficient,  $\bm{F_i}(t)$ denotes all the conservative forces acting on the $i^{th}$ monomer that includes bond and excluded-volume interactions. Further, $\bm{\Gamma}_i$ is a Gaussian white noise process with the first moment $\langle \bm{\Gamma}_i(t) \rangle = 0$  and the correlations  $\langle \bm{\Gamma}_i(t) \cdot \bm{\Gamma}_j(t') \rangle = 6 \zeta k_B T  \delta_{ij} \delta(t-t')$.\cite{risken1989fokker-planck} It describes the thermal noise, where $k_B$ is Boltzmann's constant and $T$ is the temperature. The term $F_a \bm e_i(t)$ describes the active force on monomer $i$, with the amplitude $F_a$ in the direction of the unit vector $\bm e_i$.\cite{eisenstecken2016conformational,winkler2020physics}

 We assume that the active force emerges from correlated fluctuations of the environment and describe it as colored noise. Hence, the dynamics of the orientation vector $\bm e_i$ is governed by the equation
\begin{equation}
	\dot{\bm e}_i(t) = \bm{e}_i(t) \times \bm{\xi}_i(t).
 \label{Eq:torque}
\end{equation}
Here $\bm{\xi}_i(t)$ is a Gaussian and Markovian stochastic process with the moments $\langle \bm{\xi}_i(t) \rangle = 0$ and $\langle \bm{\xi}_i(t) \cdot \bm{\xi}_j(t') \rangle = 4 D_r \delta_{ij} \delta(t - t')$. In analogy to an active Brownian particle, we denote  $D_r$ as rotational diffusion coefficient. 
The temporal correlation function of the self-propulsion velocity $\bm v_i^a = F_a \bm e_i/\zeta$ of a monomer is given by    $\langle  \bm v_i^a(t) \cdot \bm v_i^a(0) \rangle =F_a^2 \exp(-2D_rt)/\zeta^2$, i.e., $1/(2 D_r)$  is the relaxation time of the active velocity.\cite{eisenstecken2016conformational} 


Finally, the third term on the right-hand side of  Eq.~\ref{Eq:eq_motion} imposes a uniform linear shear flow along the $x$-axis (unit vector $\bm e_x$) with the gradient along the $y$-axis of the Cartesian coordinate system, with the shear rate $\dot \gamma$. Figure~\ref {Fig:schematic} schematically illustrates an active polymer along with the flow-profile gradient indicated by green arrows, and the background color depicts the gradient profile in the $y$-direction.  {\cblue  All the simulations are performed in an infinite three-dimensional bulk system since we consider an implicit solvent only.}  

{ {\it Simulation Parameters:}  In simulations, lengths are expressed in units of the LJ parameter $\sigma$, energies in units of the thermal energy $k_B T$, time in units of $\tau = \zeta \sigma^2/(k_BT) = \sigma^2/D_t$, with the translational diffusion coefficient $D_t=k_BT/\zeta$, and forces in units of $k_B T/\sigma$. The spring constant is chosen in the range, $ \kappa_s = (5 \times 10^3 - 5 \times 10^4)k_B T/\sigma^2$ to prevent stretching of the bonds in the presence of large active and shear forces. In any extreme non-equilibrium situation, the average length of the bonds has not been stretched by more than $2\%$ of the equilibrium value. The bond length is set to $l_0=\sigma$ and the LJ energy to $\epsilon/k_BT =1$.
The activity is presented by the dimensionless P\'eclet number $Pe= F_a\sigma/ (k_B T)$, which we vary in the range of $Pe = 0-300$. The ratio between the translational and rotational diffusion coefficient is set to $D_t/(D_r \sigma^2)=1/3$. 
The choice of the integration step is always in the range of $\Delta t = (10^{-7} - 10^{-4})\tau$ for the stable simulations. The flow strength is presented in terms of the dimensionless Weissenberg number $Wi_{Pe}=\dot{\gamma} \tau_r(Pe)$, where $\tau_r(Pe)$ is the longest polymer relaxation time in the presence of activity, which is indicated by the subscript $Pe$. The relaxation time is obtained from the end-to-end vector correlation 
function. The number of monomers in the polymer is taken as $N_m=200$, corresponding to the polymer contour length $L=199l_0$. For a few cases,  $N_m=400$ monomers are considered, which is indicated in the corresponding figures. 

In the ESI, we have presented results of the passive polymer for the range of $N_m=10^2-2 \times 10^3$. The translational equations of motion \eqref{Eq:eq_motion} are integrated by applying the Euler algorithm with normal distributed random displacements,\cite{ermak1978brownian}, and those of the orientation vector Eq.~\eqref{Eq:torque} are updated with the integration scheme discussed in the Ref.~\cite{winkler2015virial}.}

\section{Results}
The external flow induces large-scale conformational changes of a passive polymer, where a polymer continuously stretches and collapses in a cyclic fashion.\cite{winkler2006semiflexible,huang2010semidilute,singh2020flow} The polymer is preferentially stretched and aligned along the flow direction, while compression occurs along the gradient and vorticity directions. Additionally, in absence of shear flow, the presence of the active noise causes large-scale isotropic swelling to a flexible polymer.\cite{kaiser2014unusual,winkler2006semiflexible,anand2018structure,anand2020conformation,singh2020flow}  In this section, we elucidate the influence of the linear shear flow as well as active noise on a flexible polymer's conformational, dynamical, and rheological properties and compare the results with analytical predictions.\cite{martin2018active}

\begin{figure}[h!]
	\includegraphics[width=\columnwidth]{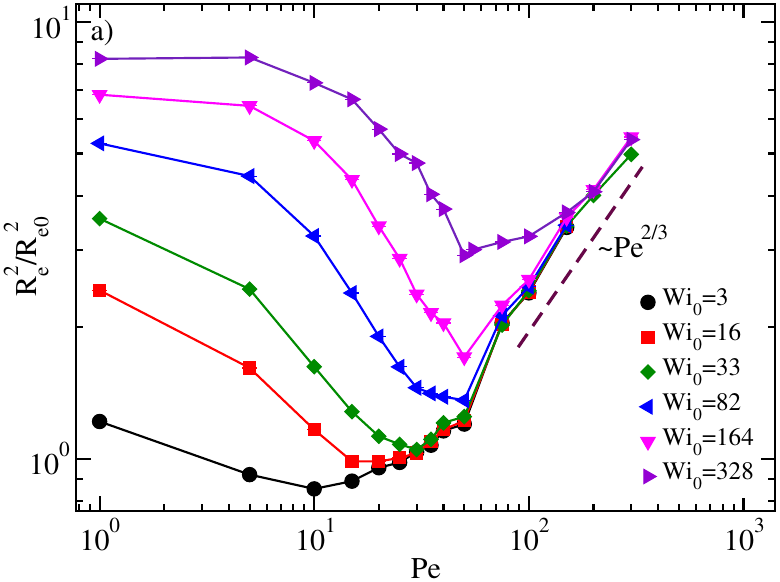}
	\includegraphics[width=\columnwidth]{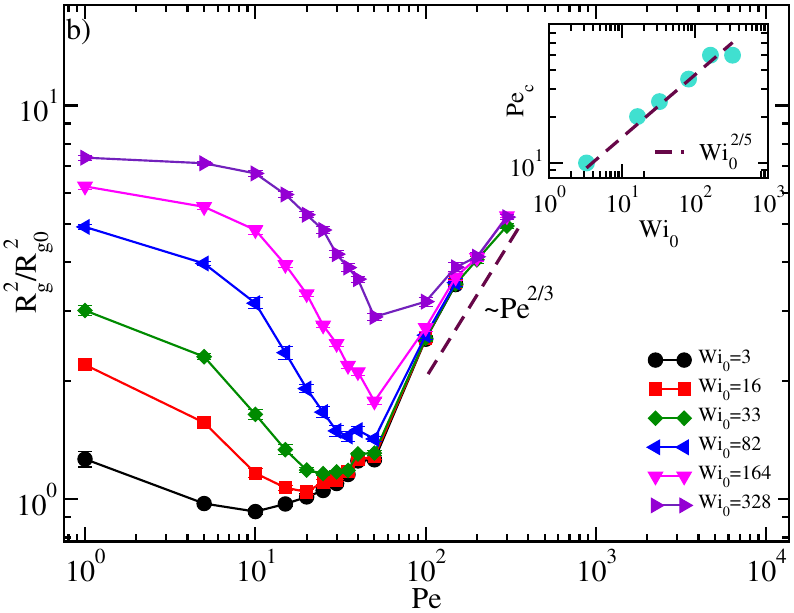} \caption{ a) Normalized mean-square end-to-end distance $R_e^2/R_{e0}^2$ as a function of the P\'eclet number $Pe$ for the  shear rates $Wi_0=3,16,33,82,164,$  and 328 at zero P\'eclet number. { Here, $R_{e0}^2$ is the mean-square end-to-end distance of the polymer at $Pe=0$  and $Wi_0=0$.}  b) Normalized mean-square radius of gyration  $R_g^2/R_{g0}^2$ as a function of $Pe$. The dashed lines show the power-law behavior of $R_e^2$ and $R_g^2$ with $Pe^{2/3}$. The inset displays the critical P{\'e}clet number at the minimum of $R_g^2$ and its power-law dependence on the Weissenberg number $Wi_0$. { Here, $R_{g0}^2$ is the mean-square end-to-end distance of the polymer at $Pe=0$  and $Wi_0=0$.} }
	\label{Fig:endtoend}
\end{figure}

\subsection{Conformations}
The conformational properties of the polymer are characterized by the mean-square end-to-end distance $R_e^2= \langle (\bm{r}_1 - \bm{r}_{N{_m}})^2 \rangle$ and the radius-of-gyration tensor
\begin{equation}
	G_{\alpha \beta} = \frac{1}{N_m}  \sum_{i=1}^{N_m} \langle \Delta{r_{i,\alpha}} \Delta{r_{i,\beta}} \rangle. 
	\label{Eq:gytensor}
\end{equation}
Here, $\alpha, \beta \in \{x, y, z\}$ and $\Delta{\bm r_i}$ is $i^{th}$ monomer's position in the polymers center-of-mass reference frame. The angular brackets indicate the average over various independent conformations of the polymer obtained during a simulation run.  Results for $R_e^2$ and $R_g^2 =\sum_{\alpha=1}^3 G_{\alpha \alpha}$ are presented in  Fig.~\ref{Fig:endtoend} as a function of $Pe$ and various shear rates $Wi_0$. 
Interestingly, both quantities exhibit a non-monotonic dependence on $Pe$, where a polymer's compression in the intermediate P\'eclet number regime is followed by stretching in the activity-dominated regime. { The change in the polymer conformations is presented relative to the equilibrium values $R^2_{e0}$ and $R^2_{g0}$, i.e., in the absence of activity and flow ($Pe=0$ and $Wi_0=0$).} In the absence of shear flow, i.e., $Wi_{Pe}=0$, the polymer swells with increasing activity, as predicted by analytical theory.\cite{eisenstecken2016conformational} In the presence of shear, $Wi_{Pe} >1$,  we can identify three distinct regimes: a weak, an intermediate, and a strong activity regime. The polymer conformations are nearly { activity independent} for weak P{\'e}clet numbers $Pe<1$. In the intermediate regime $1 \lesssim Pe  \lesssim 15$, a polymer exhibits a weak compression for $Wi_0 > 1$.  Importantly, the compression regime shifts significantly to larger P\'eclet numbers with increasing shear rate as illustrated in Fig.~\ref{Fig:endtoend} for  $Wi_0=3,16,33,82,164,$ and $328$.  After reaching the maximum compression, the polymer begins to swell with increasing  $Pe$ values -- this limit is referred to as strong $Pe$ regime --, and 
$R_e^2$ and $R_g^2$ display nearly universal behavior, where all respective curves merge into a master curve, and the polymer conformations become independent of flow strength. Here, active forces overpower shear forces. The power-law  $R_e^2 \sim   Pe^{2/3}$ agrees with the prediction of the analytical calculations for ABPOs in the absence of shear flow.\cite{eisenstecken2016conformational}. As explained in the latter article, the exponent $2/3$ is a consequence of the inextensibility of the polymer's contour length. 

The P{\'e}clet number, where the end-to-end distance attains a minimum for a given $Wi_0$, can be identified as the critical value $Pe_c$; beyond $Pe_c$, the active noise dominates over the shear flow. As presented in the inset of Fig.~\ref{Fig:endtoend}-b,  $Pe_c$ increases with $Wi_0$ according to the power law $Pe_c \sim Wi_0^{2/5}$. 

The pronounced non-monotonic behavior of active polymer conformations under shear flow is a consequence of the competition between the shear and active forces. At small $Pe<1$, shear stretches and aligns (Fig.~\ref{Fig:tan}) the polymer along the flow direction, hence $R_e^2$ increases with increasing $Wi_0$. The isotropic active forces exert an additional force, in particular transverse to the stretching direction,  and the polymer shrinks for sufficiently large $Pe$. For P\'eclet numbers above $Pe_c$, the active forces dominate over the shear forces, and the polymer swells due to the persistent motion of the ABP monomers in a uniform manner and independent of the shear rate,  with the power law $Pe^{2/3}$. \cite{eisenstecken2016conformational}
A similar effect is predicted for semiflexible active Brownian polymers in Ref.~\cite{eisenstecken2016conformational}, where active forces also imply a shrinkage of the polymer.  In this case, the isotropic active forces exert a strong force transverse to the polymer contour, which leads to shrinkage, followed by swelling as for a flexible polymer \cite{eisenstecken2016conformational}. Here, active forces dominate over bending forces.

\subsection{Radius of Gyration Tensor}

\begin{figure}[ht]
	\centering
	\includegraphics[width=\columnwidth]{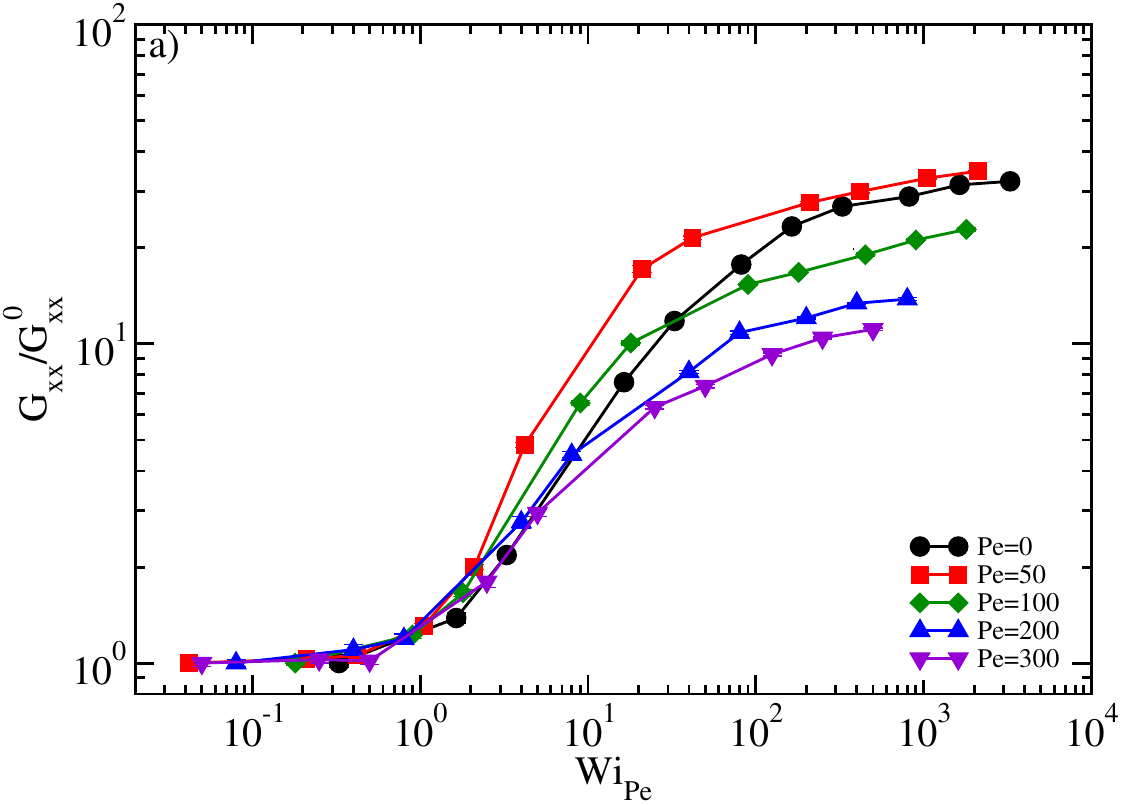}
	\includegraphics[width=\columnwidth]{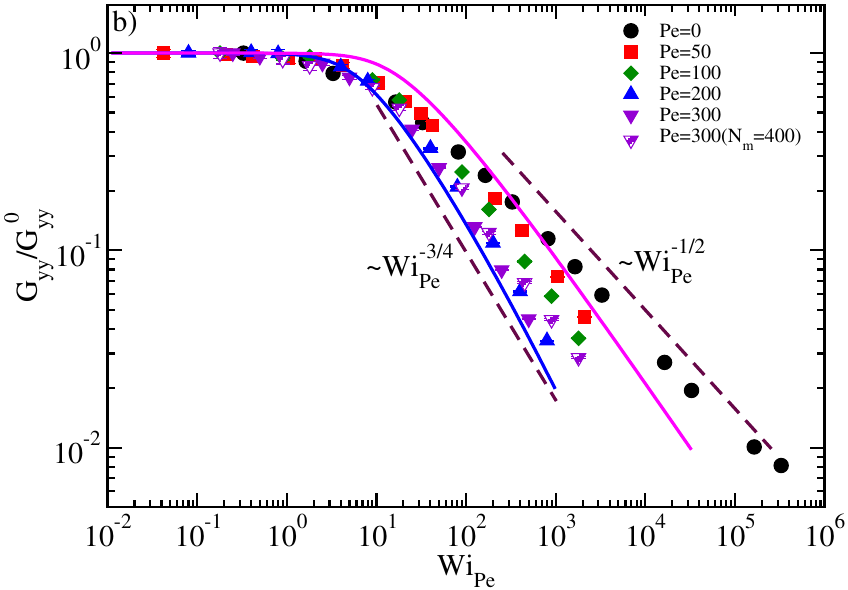}
	\caption{ a) Relative variation of the normalized gyration tensor component  $G_{xx}/G_{xx}^0$ along the flow direction and b) along the gradient direction $G_{yy}/G_{yy}^0$ as a function of $Wi_{Pe}$ for various $Pe$ and {the number of active monomers $N_m=200$}.  The dashed lines in b) indicate the power-laws  $G_{yy}/G_{yy}^0 \sim Wi_{Pe}^{-1/2}$ (right) and $G_{yy}/G_{yy}^0 \sim Wi_{Pe}^{-3/4}$ (left) for $Pe=0$ and $Pe \gg 1$, respectively. The solid lines are obtained from the analytical approach at $Pe=0$ (magenta) and $Pe=300$ (blue).\cite{martin2018active} }
	\label{Fig:gxx}
\end{figure}

\begin{figure}[t]
	\centering
	\includegraphics[width=\columnwidth]{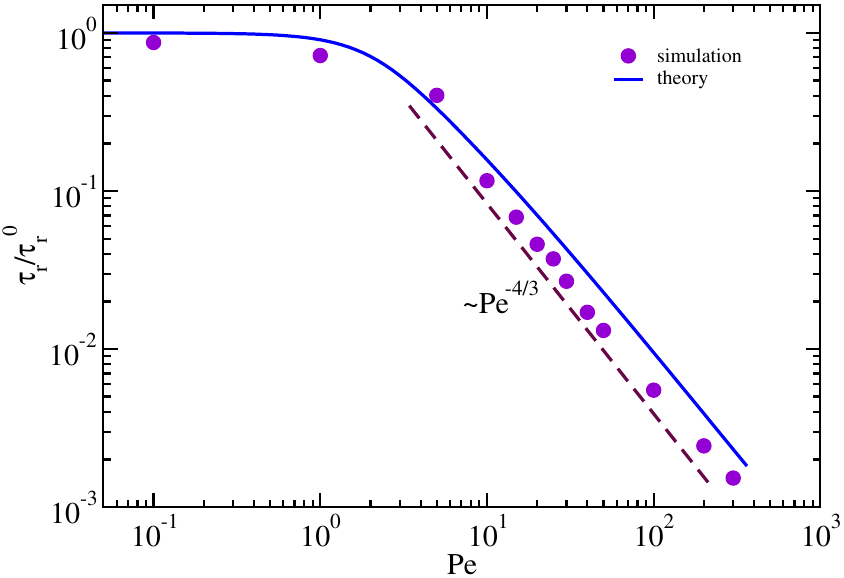}
	\caption{{Relaxation time of the active polymer end-to-end vector as a function of the P\'eclet number for the polymer length $N_m = 200$ at zero shear, $\dot \gamma =0$.} The solid blue line represents the theoretical prediction,\cite{martin2018active} while the bullets display the simulation results. The dashed line indicates the power-law $Pe^{-4/3}$, and $\tau_r^0$ is the relaxation time in the absence of activity.   }
	\label{Fig:relaxation}
\end{figure}

Figure~\ref{Fig:gxx} presents the normalized radius-of-gyration tensor component $G_{xx}/G_{xx}^0$ and $G_{yy}/G_{y}^0$ along the flow and gradient direction, respectively, as a function of shear rate for various P\'eclet numbers, where the $G_{\alpha \alpha}^{0}$ ($\alpha \in \{x,y \}$) are the gyration-tensor components in the absence of shear, but $Pe \geq 0$.  Notably, the shear strength is presented as a function of the Weissenberg number $Wi_{Pe}=\dot \gamma \tau_r(Pe)$, where $\tau_r(Pe)$ is the polymer end-to-end vector relaxation time in the presence of the active force, but in the absence of flow ($\dot \gamma=0$). { The inverse of $\tau_r(Pe)$ corresponds roughly to the onset shear rate for shear thinning}. The relaxation time is estimated from the exponential decay of the end-to-end vector correlation function  $C_{e}(t) \sim  \exp(-t/\tau_r(Pe))$.\cite{anand2018structure,singh2020flow,huang2010semidilute}  As displayed in Fig.~\ref{Fig:relaxation}, for $Pe >1$, the relaxation time  $\tau_r(Pe)$  { at zero shear} decreases with increasing  activity according to the power law $\tau_r(Pe)\sim Pe^{-4/3}$.  This result matches well with the behavior predicted by the analytical theory,\cite{eisenstecken2016conformational} in 
particular, the asymptotic scaling limit, where the longest polymer relaxation time dominates the correlation function. The latter strictly applies in the limit $N_m \gg 1$, which is not fully satisfied in the simulations.

A passive polymer acquires a coil-like shape at equilibrium. Linear polymers subjected to linear shear flow experience a force that stretches them along the flow direction, as reflected by $G_{xx}$. The steady-state results from a competition between the elastic restoring and shear forces. The shear forces stretch the polymer, and the entropic forces resist trying to keep it in the maximum entropic state, i.e.,  a coil state. As expected, the polymer monotonically elongates along the flow direction with increasing  $Wi_{Pe}$,  approaching a plateau in the larger shear limit due to the inextensibility of the polymer contour, and the polymer approaches its contour length in the limit $Wi_{Pe} \to \infty$ (Fig.~\ref{Fig:gxx}-a). 
Due to the normalization by $G_{xx}^0$, the ratio $G_{xx}/G_{xx}^{0}$ is independent of the shear rate for $Wi_{Pe} < 1$, although the polymer swells with increasing $Pe$. This swelling is reflected in the smaller plateau value at large $Pe$ because the relative variation in extension is weaker at large $Wi_{Pe}$ for larger $Pe$.

Polymer stretching along the flow direction is accompanied by a simultaneous shrinkage in the gradient and vorticity directions due to its finite contour length.  
Correspondingly, $G_{yy}/G_{yy}^{0}$ decreases with increasing Weissenberg number and exhibits a pronounced dependence on the active force, as displayed in  Fig.~\ref{Fig:gxx}-b ($G_{yy} = G_{zz})$. At low activities ($Pe \ll 1$), the shear-induced compression follows the power law $G_{yy}/G_{yy}^0 \sim Wi_{Pe}^{-1/2}$, with an exponent of nearly $1/2$ in the large shear-rate regime, as already observed in previous studies.\cite{schroeder2005characteristic,doyle1997dynamic,schroeder2018single,hur2001dynamics,shaqfeh2005dynamics,jendrejack2002stochastic,hsieh2004modeling,lyulin1999brownian,aust1999structure,colby2007shear,pincus2023dilute,tu2020direct} With increasing activity, the compression increases with increasing $Wi_{Pe}$, and in the large activity regime ($Pe \gg 1$), we obtain approximately the power law $G_{yy}/G_{yy}^0 \sim Wi_{Pe}^{-3/4}$. This indicates that the dynamics of the active polymer is different under flow in the presence of the active noise compared to that in the absence of the latter and suggests a transition from a passive to an active regime, with a significant variation in the exponent. 

Additionally, Fig.~\ref{Fig:gxx}-b shows a comparison of our simulation results with those of the theoretical approach.\cite{martin2018active} The simulation results nicely overlap with the theory for large $Pe$.\cite{martin2018active} The deviations in the limit $Pe=0$ result from the neglection of excluded-volume interactions in the analytical model. 

\subsection{Alignment}
\begin{figure}[t]
	\includegraphics[width=\columnwidth]{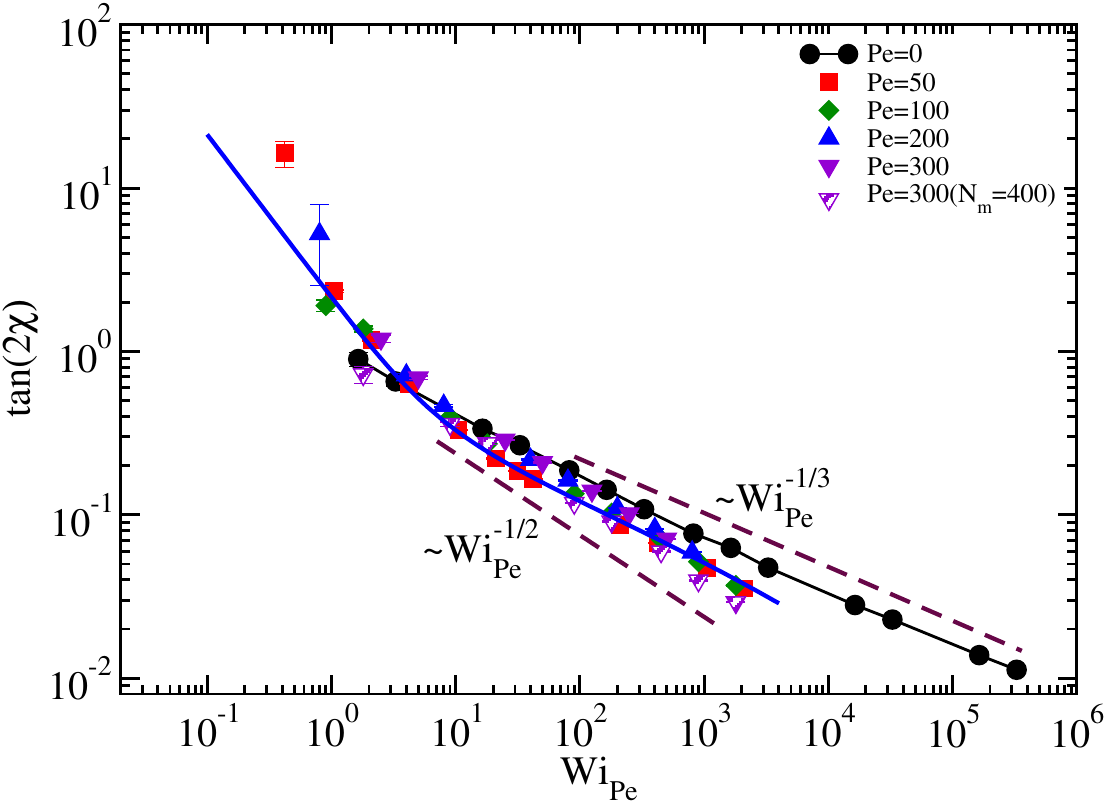}
	\caption{Polymer alignment $\tan(2\chi)$ \eqref{eq:align} as a function of the Weissenberg number $Wi_{Pe}$ for various P{\'e}clet numbers and $N_m=200$. The shaded triangles correspond to the polymer length $N_m=400$ for $Pe=300$. The dashed line (right) represents the power-law $\tan(2\chi)\sim Wi_{Pe}^{-\delta}$ with the exponent $\delta=1/3$ in the high Weissenberg regime, and the dashed line (left) that at large P\'eclet numbers with $\delta_a=1/2$. The solid line (blue) is obtained from the analytical theory at $Pe=300$   {
 the number of active monomers $N_m=200$.\cite{martin2018active}} }
	\label{Fig:tan}
\end{figure}

The shear-induced alignment of the polymer is quantified by
\begin{equation} \label{eq:align}
	\tan(2\chi) = \frac{2G_{xy}}{G_{xx}-G_{yy}}, 
\end{equation}
where $\chi$ is the angle between the eigenvector of the gyration tensor for the largest eigenvalue and the flow direction.
A passive polymer preferentially aligns in the direction of the external shear force as a consequence of the stretching of the polymer along the flow direction.
In the weak shear limit $Wi_{Pe} \to 0$, the polymer is isotropic and theory predicts  $\tan(2\chi) \sim 1/Wi_{Pe}$,\cite{winkler2006semiflexible,winkler2010conformational,martin2019active} in agreement with Fig.~\ref{Fig:tan} and simulations for passive polymers. \cite{huang2010semidilute,singh2020flow,singh2013dynamical} Here, 
thermal noise and active forces dominate over the flow.  In contrast, in the limit $Wi_{Pe} \gg 1$, the effect of flow dominates over other forces, and the polymer alignment exhibits the dependence $\tan(2\chi)\sim Wi^{-1/3}_0$ for a passive polymer (Fig.~\ref{Fig:tan}), again in agreement with theoretical predictions and previous simulations.\cite{winkler2006semiflexible,winkler2010conformational,huang2010semidilute,singh2020flow,singh2013dynamical}

The active force enhances the polymer alignment for $Wi_{Pe} \gtrsim 10$ as displayed in Fig.~\ref{Fig:tan}. The swelling of the polymer with increasing activity facilitates the stronger alignment by the flow.
In particular, the exponent of the power-law decay changes gradually, and approximately the dependence $\tan(2 \chi) \sim Wi_{Pe}^{-1/2}$ is obtained for $Pe, Wi_{Pe} \gg 1$. This is confirmed by simulations of the polymer length $N_m=400$ (Fig.~\ref{Fig:tan}). The results for the longer polymer nearly overlap with that for $N_m=200$ at the same $Pe=300$ in terms of the slope. Similar to the $G_{yy}$, we uncovered a different scaling exponent for large active forces in the limit $Pe \gg 1$ compared to a passive polymer. Again, the simulation results are in agreement with the predictions of the analytical model.\cite{martin2018active}. These observations accentuate the crucial role of activity in the shear alignment of polymers. 



\begin{figure}[t]
	\centering
	\includegraphics[width=0.95\columnwidth]{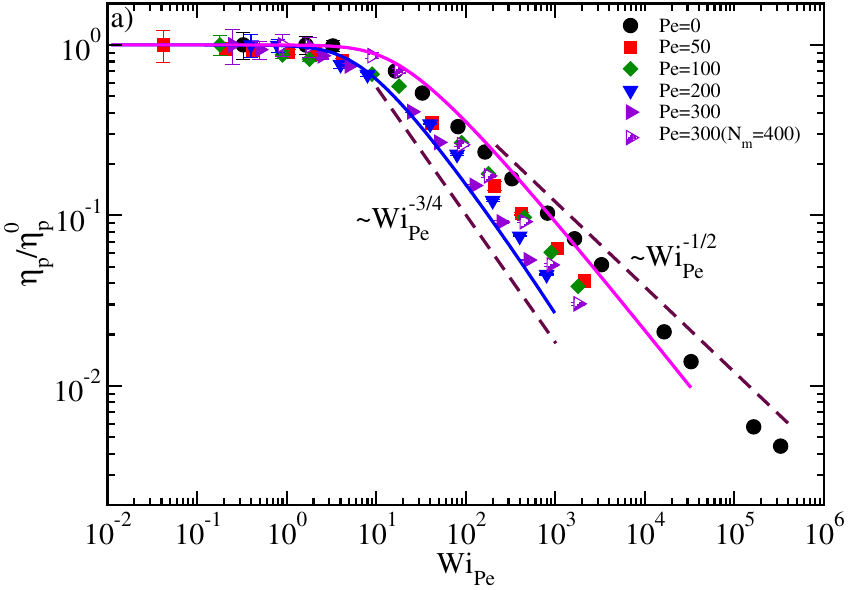}
	\includegraphics[width=0.95\columnwidth]{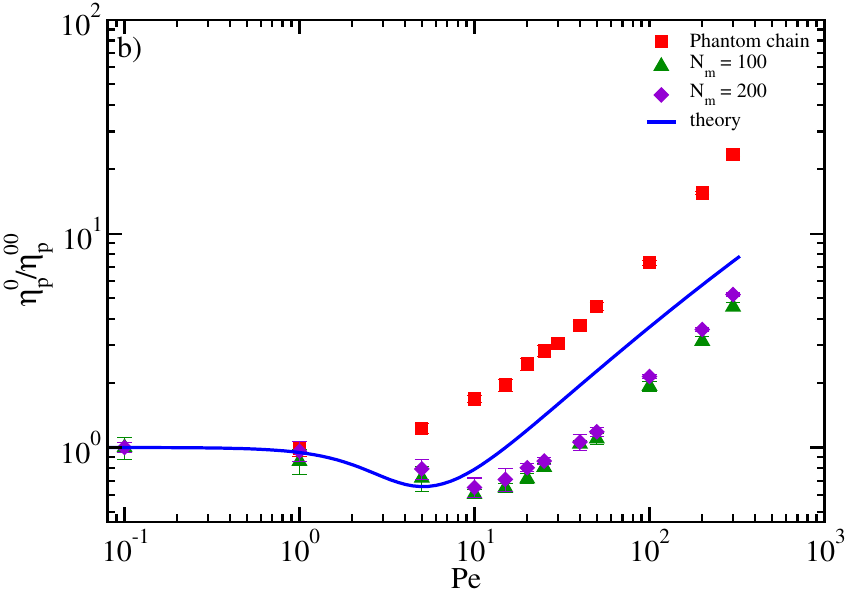}    
	\caption{a) Normalized shear viscosity $\eta_p/\eta_p^{0}$ as a function of the Weissenberg number $Wi_{Pe}$ for various $Pe$ and the number of monomers $N_m=200$ (full symbols), and $N_m=400$ and $Pe =300$ (open symbols).  The solid lines are obtained from the analytical theory for $Pe = 0$ (magenta), $Pe = 300$ (blue), and the number of active monomers $N_m = 200$.\cite{martin2018active} { The dashed lines indicate approximate power laws for $Pe=0$ (right) and $Pe \gg 1$ (left).} b) Normalized zero-shear viscosity $\eta_p^{0}/\eta_p^{00}$ as a function of the P{\'e}clet number for the polymer lengths $100$ (triangles) and 200 (diamonds), as well as the viscosity of a phantom polymer (squares) for 
 $N_m=200$. Here, $\eta_p^{00}$ is the zero shear viscosity at $Pe=0$. {The solid line (blue) represents the analytical result for a semiflexible polymer { with  $N_m = 200$ active monomers,} $pL=120$, and $\Delta = 1/3$.\cite{martin2018active}}}
	\label{Fig:viscosity}
\end{figure}

\subsection{Viscosity}
The rheological properties of the active polymer are characterized by the shear viscosity, which is calculated via shear stress. The polymer's contribution to the viscosity in the dilute limit is denoted as intrinsic viscosity. 
Using the virial expression, we determine the shear stress  according to
\begin{equation}
	\sigma_{xy} = - \left< \frac{1}{V} \sum_{i=1}^{N_m}{F}_{xi}{r}_{yi}\right> .
\end{equation}
Here, $F_{xi}$ is the total force (bond and LJ) on the $i^{th}$ monomer along the $x$-direction. The polymer's contribution to the shear viscosity is then given as $\eta_p = \sigma_{xy}/ \dot{\gamma}$. Figure~\ref{Fig:viscosity}-a displays the shear viscosity normalized by the zero-shear viscosity $\eta_p^{0}$ in the absence of shear, and Fig.~\ref{Fig:viscosity}-b presents the zero-shear viscosity itself as a function of the P\'eclet number. We like to emphasize that $\eta_p^{0}$ is a function of the active forces ($Pe$).

Two stress contributions arise from the polymer: stretching and alignment of the polymer and many-body excluded-volume interactions. 
In the weak shear limit $Wi_{Pe}<1$, the polymer behaves as in a Newtonian fluid, i.e.,  the shear stress depends linearly on the shear rate; hence, the viscosity is independent of the shear rate (Fig.~\ref{Fig:viscosity}-a). At higher values of the shear rate ($Wi_{Pe}>1$), the polymer exhibits shear thinning, and $\eta_p$ decreases with increasing $Wi_{Pe}$. This decrease attributes to the alignment and the stretching of the linear polymer. In the asymptotic limit $Wi_{Pe} \gg 1$, the shear thinning behavior can be described by a power law as $\eta_p \sim Wi_{Pe}^{-\beta}$. As is evident from Fig.~\ref{Fig:viscosity}-a, the exponent  $\beta$ depends on the activity $Pe$. For a passive polymer, we find $\beta = 1/2$, in agreement with previous simulations.  \cite{schroeder2018single,larson2005rheology,pan2014universal,huang2010semidilute,liu2004brownian} To ascertain the scaling laws for $\eta_p$ in the asymptotic limits, we have performed simulations for the longer polymers with $N_m=400$ in the limit $Wi_{Pe} > 10^2$ for $Pe=0$ and $Pe=300$. Here, we confirm the power-law exponent $\beta=1/2$ for passive polymers. The theoretical approach predicts a stronger drop of $\eta$ for $Wi_{Pe} \gg 1$, which is related to the lack of excluded-volume interactions in that approach. 

Activity severely influences the intrinsic viscosity of the polymer. Figure~\ref{Fig:viscosity}-a shows that the shear-thinning behavior becomes more prominent with increasing $Pe$. In the limit of $Pe \gg 1$, a power-law regime seems to be present for $Pe>200$ and $10 < Wi_{Pe} <10^3$, with the exponent $\beta =3/4$. The different scaling exponents reflect the differences in polymer conformations in the presence and absence of activity. The active forces facilitate faster and stronger stretching and alignment, which implies a more pronounced viscosity change. A comparison of our simulation results with the analytical approach shows very good agreement. The analytical model predicts the asymptotic behavior $\eta_p \sim Wi_{Pe}^{-1}$ in the limits $Pe \gg 1$ and $Wi_{Pe} \to \infty$. The good agreement between simulation results and theory suggests that the presented range of $Wi_{Pe}$ corresponds to a crossover regime, and the asymptotic behavior is only reached for significantly larger shear rates.

Figure~\ref{Fig:viscosity}-b presents the dependence of the zero-shear viscosity ($\eta_p^{0}$) on the active force and reflects its dependence on the excluded-volume interactions, which imply a non-monotonic dependence on $Pe$. The zero-shear viscosity of a phantom polymer increases monotonically with increasing P\'eclet number for $Pe>1$, consistent with theoretical predictions.\cite{martin2019active} However, $\eta_p^0$ of a excluded-volume polymer decreases with increasing $Pe$ in the range $1< Pe<15$. For large $Pe$, the zero-shear viscosity grows monotonically again, similar to that of a phantom polymer, where $\eta_p^0$ seems to be polymer-length independent. 
The drop in $\eta_p^0$ is associated with the conformational changes of the polymer in the intermediate activity region ($1<Pe<15$) (see Fig.~\ref{Fig:endtoend}), where excluded-volume interactions cause a polymer shrinkage. \cite{anand2019behavior,das2021coil,eisenstecken2017conformational}  

{ The influence of polymer shrinkage is illustrated by a comparison with analytical calculations of the zero-shear viscosity of a semiflexible active Brownian polymer.\cite{martin2018active} Such polymers also shrink initially with increasing P\'eclet number, before activity causes swelling.\cite{eisenstecken2016conformational}  The random active forces imply enhanced fluctuations transverse to the backbone of the semiflexible polymers, which results in polymer shrinkage. {In the case of flexible polymers with excluded-volume interactions, the active force weakening the latter interactions.} The comparison of the zero-shear viscosity in Fig.~\ref{Fig:viscosity}-b from simulations and theory of semiflexible polymers demonstrates the reduction in $\eta_p^0$ by their shrinkage.  The parameters of the semiflexible polymer are $pL = 120$, $N_m= 200$, and $\Delta=D_t/(l_0^2 D_r) = 1/3$. This corresponds to a reduction of the radius of gyration to approximately $80\%$  of the equilibrium value of the semiflexible polymer in the absence of the flow,  similar to the activity-driven shrinkage of the excluded-volume polymer observed in simulations in the absence of the flow.\cite{anand2020conformation}  Here, $pL$ represents the ratio of polymer length ($L$) and the persistence length   $l_p=1/(2p)$  of the semiflexible polymer.}

\section{Conclusion}

In this article, we have presented results of coarse-grained over-damped simulations for the conformational, dynamical, and rheological properties of an active Brownian polymer under linear shear flow. 
In particular, the interplay of activity and flow is highlighted, which leads to the complex coupling of a polymer's structural and dynamical properties.  We have addressed the influence of the non-equilibrium forces: linear shear flow, which always aligns the polymer along the flow, and active forces, which stretch a polymer.  The stretched polymer under shear flow displays a non-monotonic conformational response {as a function of active noise. The stretching along the shear direction is reduced by active forces, causing shrinkage of the polymer, and with a further increase of the active force beyond a critical P{\'e}clet number ($Pe_c$), its reswelling.} The obtained $Pe_c$ grows with the Weisenberg number as $Pe_c  \sim Wi_0^{2/5}$, demonstrating that $Pe_c$, as expected, increases with the strength of the active noise.   

Furthermore, we have demonstrated that various physical properties such as the radius-of-gyration tensor component $G_{yy}$, alignment $\tan(2\chi)$,  and shear viscosity $\eta_p$, exhibit universal behavior independent of $Pe$ in the small shear rate regime $Pe<15$ when these quantities are presented as a function of the dimensionless Weissenberg number $Wi_{Pe}=\dot{\gamma} \tau_r(Pe)$, with the active force-dependent relaxation time $\tau_r(Pe)$. At large shear rates $Wi_{Pe} \gg 1$, scaling relations at small active forces are the same as those of passive polymers despite the presence of the active noise; in particular, we find for the shear viscosity the power-law exponents $1/2$, for the alignment $1/3$, and for the shrinkage of the polymer in the gradient direction $1/2$, in agreement with previous simulation studies. \cite{schroeder2005characteristic,schroeder2018single,li2000comparison,shaqfeh2005dynamics} These exponents are verified for longer exclude-volume polymers with $N_m=800$ and $N_m=2000$ (here, also a phantom polymer has been studied); see ESM.  
In the limit of large activities, the scaling relations change significantly. Specifically, we approximately find $\tan(2\chi)\sim Wi_{Pe}^{-1/2}$, $G_{yy} \sim Wi_{Pe}^{-3/4}$, and  $\eta_p \sim Wi_{Pe}^{-3/4}$, where the latter power-law may correspond to a cross-over regime rather than the asymptotic for $Wi_{Pe} \to \infty$. 
These changes indicate a pronounced effect of activity on polymer conformations and alignment and, as a consequence, on the rheological properties. The effect of the active noise on the polymer conformations can also be visualized in the supporting Movies for various shear rates at the given P\'eclet number $Pe=50$, see (ESI-Movie-1, 2, and 3).  The comparison of the simulation results for $Pe \gg 1$ with those of the analytical approach of Ref.~\cite{martin2018active} shows good agreement. Certain scaling exponents obtained in our simulations differ slightly from those predicted by theory. This may be due to limitations of the simulations, which have not yet reached the asymptotic limit of $Pe\rightarrow\infty$ and $Wi_{Pe}\rightarrow  \infty$, or the absence of excluded-volume interactions in the analytical model.  

{ Remarkably, excluded-volume interactions lead to a decrease in the zero-shear viscosity over a certain range of P\'eclet numbers,  followed by a viscosity increase for large $Pe$.} This decrease is a consequence of the compression of the polymer by a weakening of the excluded-volume interactions due to the active forces.\cite{anand2020conformation} Here, the presence of excluded-volume interactions is most pronounced, whereas it only weakly affects 
the polymer conformational and rheological properties for large activities and shear rates.

In summary, our results highlight the importance of active noise and excluded-volume interactions on polymers in shear flow. We demonstrate that the contribution of an active polymer to the solution viscosity can decrease in the presence of active forces, and scaling properties of various conformational and dynamical quantities also change. In particular, the role of excluded-volume interactions is crucial in an intermediate activity regime.\cite{anand2018structure,das2021coil} 
The activity-enhanced shear thinning suggests that active or activated polymers can be used as viscosity modifiers, as has also been shown for bacterial solutions.\cite{lopez2015turning}
The active forces lead to distinct physical behaviors compared to thermal noise. The effect of polymer-polymer interactions, specifically in semidilute regimes and the effect of semiflexibility, would unravel further particular rheological features of such systems. Incorporating these features would take our model closer to more realistic biological systems.\cite{deblais2020rheology} Such simulation results may be comparable with experimental results from the active worms,\cite{deblais2020rheology} flexible polymers suspended in the solution of active colloids.\cite{bechinger2016active,palacci2013living} 
 
{\cblue Our simulations lack hydrodynamic interactions (HI) for the model's simplicity and as a reference to a dry system under the influence of Gaussian colored noise. 
As has been demonstrated by simulations and theory,\cite{winkler2020physics,martin2019active,mart:20} HI can change the conformational and dynamical properties of ABPOs substantially, depending on the nature of the active force. Most pronounced changes are obtained for self-propelled monomers, whereas externally driven monomers exhibit more similar conformational and dynamical features as polymers in the absence of HI.\cite{winkler2020physics}  Here, further simulations are certainly desirable to resolve the effect of shear flow on the rheological properties of such polymers, specifically to resolve the influence of the propelling mechanism. }

\section{Acknowledgment}
SPS  acknowledges financial support from the DST-SERB Grant No. CRG/2020/000661. The computational facilities at IISER Bhopal and Paramshivay NSM facility at IIT-BHU are highly acknowledged for providing computational time.

\balance


\bibliography{Main,bibliography} 
\bibliographystyle{rsc} 
\include{temp1}
 \end{document}

%% file: temp1.tex





\title{Supplementary Material: Characteristic  features of self-avoiding active Brownian polymers under linear shear flow}
\thanks{A footnote to the article title}%



\author{Arindam Panda, Roland G. Winkler, Sunil P. Singh }


\date{}

\maketitle

\begin{figure}[ht]
\centering
	\includegraphics[width=\linewidth]{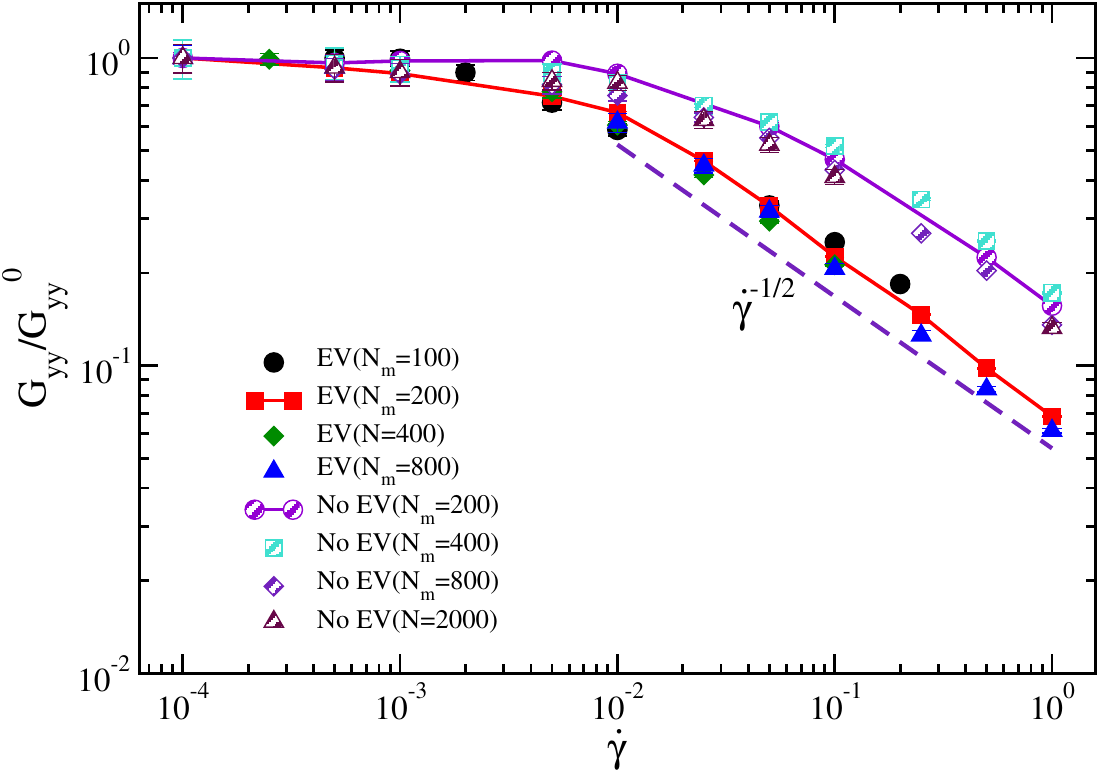}
	\caption{The variation of normalized values of $G_{yy}/G_{yy}^0$ of a passive polymer as a function of shear rate($\dot \gamma$) for different polymer lengths ranging from $N_m = 100 - 800$ with EV and $N_m = 200 - 2000$ without EV (phantom polymer). The dash line indicates the power law variation of $G_{yy}/G_{yy}^0 \sim \dot \gamma^{-1/2}$, with an exponent $-1/2$.}
	\label{Fig:gyy_passive}
\end{figure}
\begin{figure}[ht]
\centering
	\includegraphics[width=\linewidth]{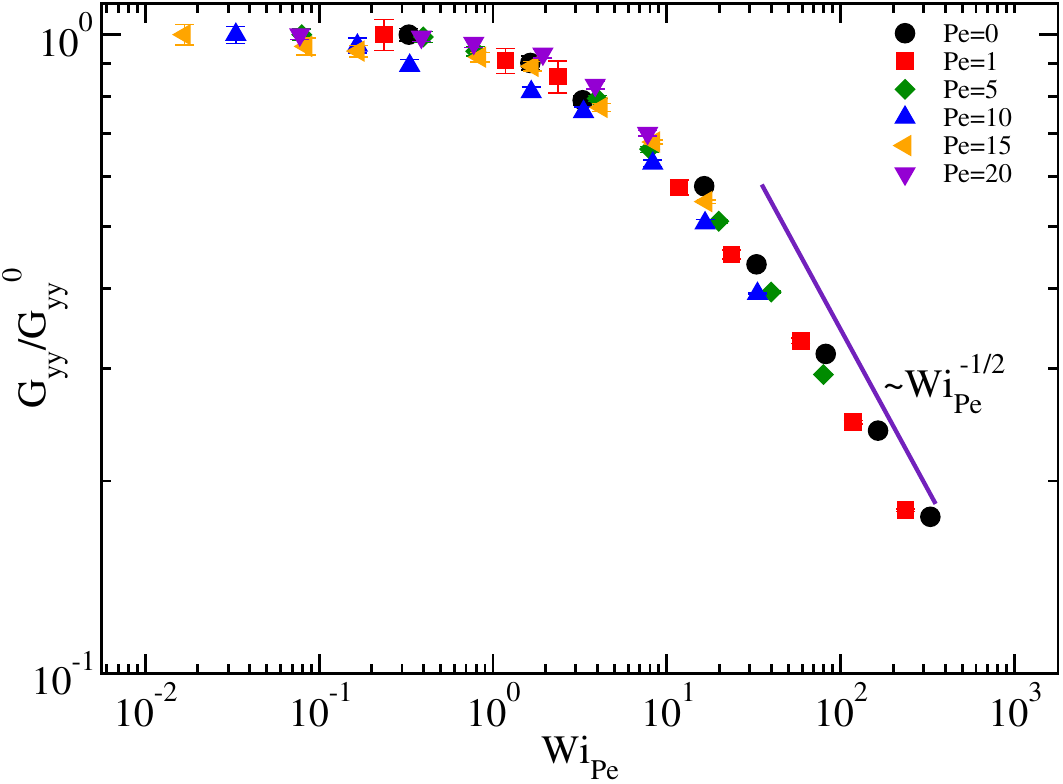}
	\caption{The variation of $G_{yy}/G_{yy}^0$  for various $Pe$ as a function of $Wi_{Pe}$. The solid line indicates the power law variation as $G_{yy}/G_{yy}^0 \sim Wi_{Pe}^{-1/2}$, with an exponent $1/2$.}
	\label{Fig:gyy}
\end{figure}
\subsection{Radius of gyration:}
We performed extensive  simulations to analyze the behavior of passive polymers of various lengths, both with and without excluded volume (EV) interactions;
specifically, we examine for polymer lengths $N_m = 100, 200, 400,$ and $800$ with EV, and $N_m = 200, 400, 800,$ and $2000$ without EV. In both cases, we observed a scaling exponent of $1/2$ for the compression of the polymer along the gradient direction. Notably, in the asymptotic limit $G_{yy}/G_{yy}^0$ decreases according to the power law $G_{yy} \sim \dot \gamma^{-1/2}$ and remains the same regardless of real or phantom polymers, as depicted in Fig.~\ref{Fig:gyy_passive}. 

Additionally, we also present the variation of the polymer compression in the gradient direction in the intermediate regime of the P\'eclet numbers $Pe=1,5,10, 15$ and $20$.  
In this limit, the variation of $G_{yy}/G_{yy}^0$  as a function of $Wi_{Pe}$ for all the curves tend to converge and overlap with each other, displaying a universal behavior,  see Fig.~\ref{Fig:gyy}. This convergence suggests a minimal role of the active noise in this limit. The shear dominates over the active noise. 
Furthermore, a solid line depicts the power-law behavior in the graph, indicating a relationship described by $G_{yy}/G_{yy}^0 \sim Wi_{Pe}^{-1/2}$ with an exponent of $1/2$ in the large shear regime.


\begin{figure}[t]
	\includegraphics[width=\columnwidth]{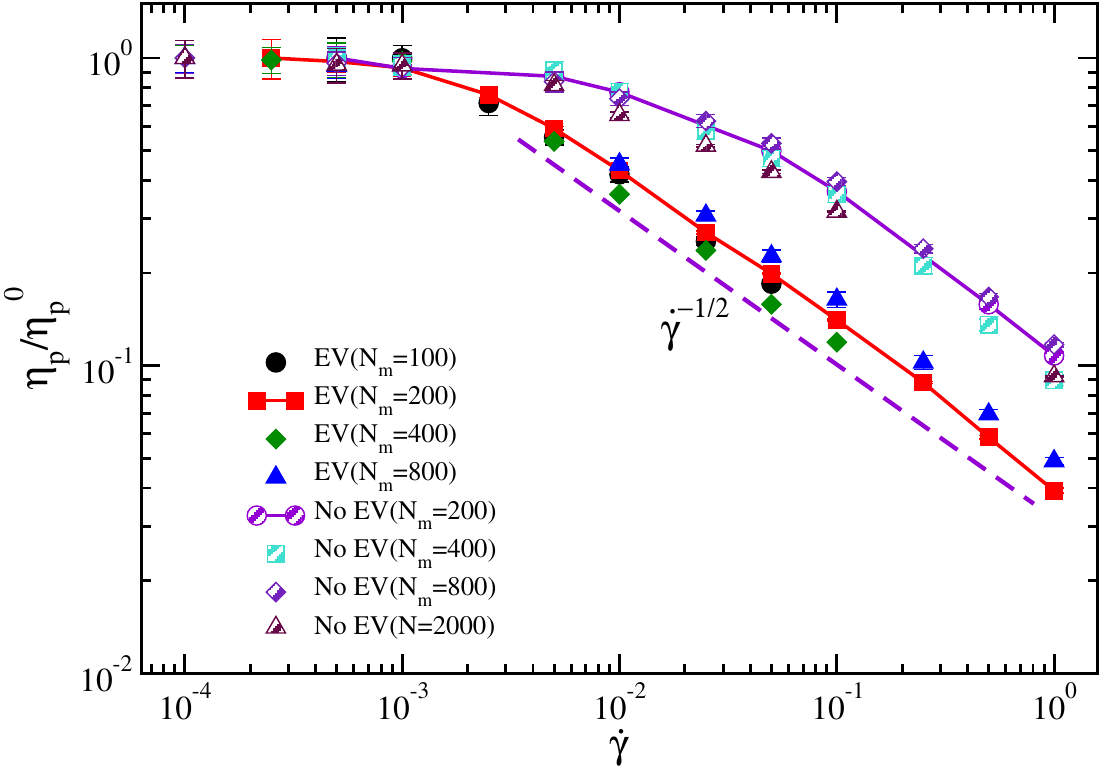}
	\caption{ The normalized shear viscosity $\eta_p/\eta_p^{0}$ of a passive polymer as a function of shear rate($\dot \gamma$) for different polymer lengths ranging from $N_m = 100 - 800$ with EV and $N_m = 200 - 2000$ without EV (phantom polymer). The dashed line displays the power law variation of the viscosity as $\eta_p/\eta_p^0 \sim \dot \gamma^{-1/2}$, with an exponent $1/2$.}
\label{Fig:viscosity_pass}
\end{figure}

\begin{figure}[t]
\includegraphics[width=\columnwidth]{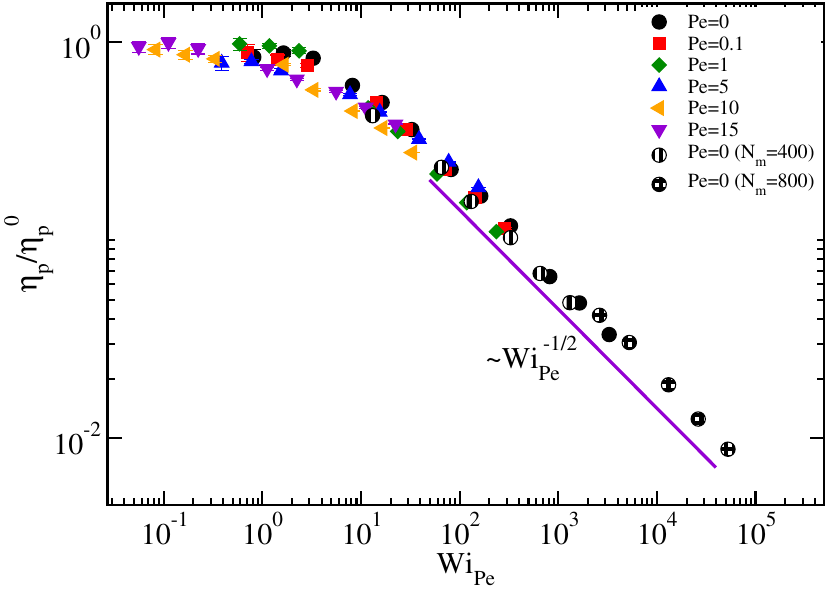}
	\caption{ The normalized shear viscosity $\eta_p/\eta_p^{0}$ as a function of P{\'e}clet number dependent Weissenberg number ($Wi_{Pe}$) for various  P\'eclet number $Pe$ (in the intermediate regime of Pe). The shaded symbols display viscosity for polymer lengths $N_m=400$ and $800$.}
	\label{Fig:viscosity}
\end{figure}

\newpage
\subsection{Shear Viscosity}
Now, we present the variation of the viscosity of the polymer for the longer polymers. The polymer lengths considered are $N_m=100,200,400$ and $800$ for the EV, and without EV (phantom), the polymer lengths are given as $N_m=200,400,800$, and $2000$. In both cases, the shear viscosity in the asymptotic limit of the shear rates decreases according to the power law $\eta_p\sim \dot \gamma ^{-1/2}$ similar to  $G_{yy}$.  The exponent of the real and phantom polymers are nearly identical, as indicated in  Fig.~\ref{Fig:viscosity_pass}. To distinguish the variation in the viscosity of the phantom and real polymers, they are plotted as a function of the shear rate ($\dot \gamma$). 

The intrinsic viscosity of the polymer as a function of $Wi_{Pe}$ in the intermediate regime of the $Pe$ is displayed in Fig.\ref{Fig:viscosity}. In this regime, normalized viscosity can be described by a universal curve with $Wi_{Pe}$ for the $Pe=0.1,1,5,10,15$, suggesting that the role of the active noise in this regime accounts for the faster relaxation of the polymer. Therefore, the effect noise can be accounted for by renormalizing the Weissenberg number $Wi_{Pe}$. \\

\subsection{Supporting movies}
The movie files demonstrate the active Brownian polymer's conformational dynamics subjected to linear shear flow. \\

ESI-Movie S1: The movie reveals the polymer's conformational dynamics for $N_m = 200$, $Wi_{Pe} = 0.04$ and $Pe = 50$. During this process, the polymer undergoes stretching in all directions. \\ 

ESI-Movie S2: 
The movie reveals the polymer's conformational dynamics for $N_m = 200$ in the presence of active noise,  at Weissenberg number $Wi_{Pe} = 4.0$ and $Pe = 50$.   \\ 

ESI-Movie S3: 
The movie reveals the polymer's conformational dynamics for $N_m = 200$ in the presence of active noise at high shear rates given as Weissenberg number $Wi_{Pe} = 42$ and $Pe = 50$. The polymer is sufficiently stretched in the shear direction despite the strong active noise.



%